\documentclass{article}

\usepackage[dvips]{graphics}

\begin{document}

\title{The role of network topology on extremism propagation with the Relative Agreement opinion dynamics}

\author{Fr\'ed\'eric Amblard, Guillaume Deffuant}

\maketitle

\begin{center}
Laboratoire d'Ing\a'enierie des Syst\a`emes Complexes\\
 Cemagref\\
    24, avenue des Landais \\
    63172 Aubi\`ere Cedex, France.
\end{center}

%%%%%%%%%%%%%%%%%%%%%%%%%%%%%%%%%%%
\begin{abstract}
In \cite{Deffuantetal2002}, we proposed a simple model of opinion
dynamics, which we used to simulate the influence of extremists in
a population. Simulations were run without any specific
interaction structure and varying the simulation parameters, we
observed different attractors such as predominance of centrism or
of extremism. We even observed in certain conditions, that the
whole population drifts to one extreme of the opinion, even if
initially there are an equal number of extremists at each extreme
of the opinion axis. In the present paper, we study the influence
of the social networks on the presence of such a dynamical
behavior. In particular, we use small-world networks with variable
connectivity and randomness of the connections. We find that the
drift to a single extreme appears only beyond a critical level of
connectivity, which decreases when the randomness increases.
\end{abstract}

%%%%%%%%%%%%%%%%%%%%%%%%%%%%%%%%%%%%%%%%%%%%%%%%%%%%%%%%%%%%%%%%

\section{Introduction}

Most of the social influence literature, especially the
contribution from social scientists, insists on the role of
pre-existing social structures, such as social networks. In general
we lack empirical data about the topology of social network,
although recent efforts lead to the proposal of different models
such as small-worlds \cite{WattsandStrogatz1998} and scale-free networks \cite{BarabasiandAlbert2002,Barabasi2002}. The purpose of this
paper is to study the influence of the network structure on
opinion dynamics. 
Most of the models of opinion formation in the litterature are based on an Ising-like influence dynamics of binary or discrete opinions \cite{WeidlichHagg1983,Weidlich1991,Galam1997} taking as an assumption the influence of the group as the whole on the individual opinion. Another well-studied model is the one proposed by Sznajd and its numerous extensions \cite{Sznajd2000}. The influence of network topology, especially of small-world networks, on such models have been studied \cite{BarratWeigt2000,KupermanZanette2002}
In \cite{Deffuantetal2002} we proposed an
individual-based simulation model of continuous opinion dynamics, the
relative agreement model (RA model). This model can be considered
as a variant of the "bounded confidence" model
\cite{HegselmannetKrause2002,Deffuantetal2001,Weisbuchetal2002a,Weisbuchetal2002b}
based on random pair-interaction in which an individual influences another one only if the distance
between their opinions, considered as continuous, is below a given threshold. Related models with a vector of binary traits rather than continuous opinions can be found in the litterature \cite{Axelrod1997,Weisbuchetal2002a}.
In the
"classical" bounded confidence model, the influence varies
linearly with the distance between the opinions, until the
threshold where it suddenly goes to zero. The Relative agreement
model considers opinions segments, defined by the opinion and an
uncertainty around this opinion. Both the opinion and the
uncertainty are real numbers. The influence takes into account the
overlap between the two opinions segments and avoids the
discontinuity. Moreover, the more certain is the agent, the more
convincing or influential it is. We model extremists in the
population as individuals with a very low uncertainty and an
opinion located at the extremes of the initial opinion
distribution. When such extremists are initially present in the
population, and when each individual is connected to all the
others, the simulations exhibit three types of convergence: the
"central convergence" in which the extremists attract only a very
limited part of the population, which is initially close to the
extremes; the "both extreme convergence" in which the population
splits into two groups, each converging to one extreme; the
"single extreme convergence" in which the whole population
converges to one of the extremes \cite{Deffuantetal2002}. The last type of convergence was
the most unexpected. It appears even when the initial number of
extremists at each extreme is the same. In this paper, we study
the influence of some social network structures on the occurrence
of the single extreme convergence.

In a first paragraph we briefly present the Relative Agreement
Model with extremists and the main results for the fully connected
case. Readers who wanted to know more about this model can refer to \cite{Deffuantetal2002}. 
We then consider social networks defined on a regular
lattice with a Moore neighborhood. The single extreme convergence
does not occur with these social networks. We then explore the
behavior of the model on Small-World networks \cite{WattsandStrogatz1998} that enable us to
tune both the average connectivity of the network and the
randomness of its connections. The simulations show that there is
a critical level of connectivity beyond which the drift to a
single extreme appears. We then discuss the observed behavior of
our model.

%%%%%%%%%%%%%%%%%%%%%%%%%%%%%%%%%%%%%%%%%%%%%%%%%%%%%%%%%%%%%%%%%%%%%%%%%%%%%%%%
\section{The Relative Agreement Model}

\subsection{Definition}

We consider a population of $N$ individuals. Each individual
$i$ is characterized by two variables, its opinion
$x_i$ and its uncertainty $u_i$, both being real
numbers. We call segment $s_i=[x_i-u_i, x_i+u_i]$ of the
opinion axis the opinion segment of individual $i$. In the
following, we draw the opinions from a uniform distribution
between -1 and +1. We suppose that random pair interactions take
place among the individuals, during which they may influence each
other's opinion segment. In the simplest model of "bounded
confidence" another individual $j$ can influence agent $i$ whenever
its opinion is inside segment $s_i$. We here use the Relative
Agreement model which results in a continuous variation of the
influence on the opinion axis and where the intensity of the
influence depends on the agent $j$ uncertainty $u_j$. More precisely,
let us consider opinion segments $s_i=[x_i-u_i, x_i+u_i]$ and $s_j=[x_j-u_j, x_j+u_j]$. We define the agreement of
agent $i$ with $j$ (it is not symmetric) as the overlap of $s_i$ and $s_j$,
minus the non-overlapping part.
The overlap $h_{ij}$ is given by:
\begin{equation}
h_{ij}=min(x_i+u_i,x_j+u_j)-max(x_i-u_i,x_j-u_j)
\end{equation}

The non-overlapping width is:
\begin{equation}
2.u_i-h_{ij}
\end{equation}

The agreement is the overlap minus the non-overlap:
\begin{equation}
h_{ij}-(2.u_i-h_{ij})=2.(h_{ij}-u_i)
\end{equation}

The relative agreement is the agreement divided by the length of
segment $s_i$:
\begin{equation}
\frac{2.(h_{ij}-u_i)}{2.u_i}=\frac{h_{ij}}{u_i}-1
\end{equation}

If $h_{ij}>u_i$, then the modifications of $x_j$ and $u_j$ by the interaction with
$i$ are multiplied by the relative agreement:
\begin{eqnarray}
x_j:=x_j+\mu.(\frac{h_{ij}}{u_i}-1).(x_i-x_j)\\
u_j:=u_j+\mu.(\frac{h_{ij}}{u_i}-1).(u_i-u_j)
\end{eqnarray}

Where $\mu$ is a rate of the dynamics.
If $h_{ij}\leq u_i$, there is no influence of
$i$ on $j$. The main features of the relative agreement model are:
\begin{itemize}
\item During interactions, agents not only influence each other's
opinions but also each other's uncertainties.
\item The influence is not symmetric when the agents have different uncertainties;
"confident" agents (low uncertainty) are more influential.
\end{itemize}
The influence (the modifications of $x_j$ and $u_j$) varies continuously
when $x_j$, $u_j$, $x_i$ and $u_i$ vary continuously.
Then, as a reminder, the only stochastic part of this model concerns the selection of the pair-interactions that are drawn at random.

\subsection{Addition of extremists}
We now introduce extremists into our population: we suppose that these individuals have opinions located at the extremes of the opinion distribution are more confident than the non-extremists (moderate) individuals. We define therefore two values, on the one hand, their initial uncertainty: $u_e$ the uncertainty of all the extremists, and on the other hand, $U$ the initial uncertainty of the moderate, supposed higher than $u_e$.
We define also $p_e$ as the global proportion of extremists in the population. $p_+$ and $p_-$ are then the proportion of extremists at the positive or negative extreme opinions.
The relative difference between the proportion of positive and negative extremists is noted $\delta$:
\begin{equation}
\delta =\frac{\mid p_+-p_- \mid}{p_++p_-}
\end{equation}

In practice, we first randomly draw opinions of $(1-p_e).N$ agents of the population from a uniform distribution between -1 and +1. Then we initialize $Np_+$ agents to +1 and $Np_-$ most negative opinions to -1, moreover we initialize them with the uncertainty $u_e$, and the others with the uncertainty $U$.

\subsection{Results on the totally connected case}

\subsubsection{Three convergence types}
From \cite{Deffuantetal2002}, this very simplified model of extremism, exhibits 3 different dynamical regimes, depending on the parameters: the "central convergence" in which the extremists have a small influence on the rest of the population; the "both extremes convergence" in which the whole population becomes extremist with an almost equal number of extremists at each extremes; and the "single extreme convergence" where the whole population drifts to one single extreme.
The following set of figures, obtained from numerical simulations, exhibits the three different dynamical regimes. The $x$-axis codes for time (number of iterations), the $y$-axis for opinions, and the level of gray for uncertainty. Each trajectory allows following the evolution in opinion and uncertainty of one individual agent. Common parameters are, $\mu =0.5, \delta =0, u_e=0.1, N=200$. The uncertainty parameter $U$ of the general population is increased from Fig.1(a) to Fig.1(d).
Fig. 1(a), obtained for $U = 0.4$, shows an example of central convergence. In this case, only a marginal part of the initially non-extremists became extremist (4\%). Fig. 1(b), obtained for $U = 1.2$ shows an example of bipolarization. In this case, the moderate agents are attracted by one of the extremes according to their initial position. Fig. 1(c) obtained for $U = 1.4$ shows an example of single polarization. In this case, the majority of the population is attracted by one of the extremes. This behavior can take place even when the number of initial extremists is the same at both extremes. But for another sample drawn from the same initial distribution, all other parameters being equals, one can even observe a central convergence  (see Fig. 1(d)). The sensitivity to initial conditions and to random sampling is a general feature of the model in transition regions.

\begin{center}
Fig.1
\end{center}

The instability of attractor is confirmed by the master equation analysis \cite{Faureetal2003}. For a perfect initial uniform distribution of opinions, the master equation (which is deterministic) displays a symmetric attractor (either central or on both extremes) in the region where single extreme convergence is obtained with the multi-agent RA model. But any slight asymmetry in the initial distribution changes the central convergence into a single sided extreme attractor (which side being determined by the asymmetry).
The coding of the uncertainties shows that in all three cases clustering not only occurs among the opinions but also among uncertainties: e.g. when extremism prevails, it prevails in both opinions and uncertainties.

\subsubsection{General results of the parameter space exploration}
\paragraph{Convergence type indicator}
We expressed the results of the exploration with an indicator of the convergence type, denoted $y$. To compute indicator $y$, we consider the population of opinions after convergence, and:

\begin{itemize}
\item We compute the proportions $p'_+$ and $p'_-$ of the initially moderate agents which became extremists in the positive extreme or negative extreme.
\item   The indicator is then defined by: 
\begin{equation}
y= p_+^{'2} + p_-^{'2}
\end{equation}

\end{itemize}

The value of this indicator indicates the type of convergence:
\begin{itemize}
\item   If none of the moderate agents becomes extremist, then $p'_+$ and $p'_-$ are null and $y = 0$.
\item   If half of the initially moderate converge to the positive extreme and half to the negative one, we have $p'_+ = 0.5$ and $p'_- = 0.5$, and therefore $y = 0.5$.
\item   If all the moderate agents go to only one extreme (say the positive one), we have $p'_+ =1$ and $p'_- = 0$, and therefore $y = 1$.
\item   The intermediate values of the indicator correspond to intermediate situations.
\end{itemize}

\paragraph{Typical patterns of $y$}
We found that the exploration of the model can be conveniently presented as variations of $y$ with $U$ and $p_e$. This representation leads to one typical pattern of average $y$ for $\delta = 0$, and another one for $\delta > 0$. There is therefore a significant change between the cases where the proportion of positive and negative extremists is exactly the same, and when it is slightly different. When the other parameters $u_e$ and $\mu$ are modified, the global shape of the patterns remains similar: only the position of the boundaries between the convergence zones varies.

\begin{center}
Fig.2
\end{center}

The typical patterns obtained for $\delta = 0$ and $\delta > 0$ are shown on Fig. 2. In this figure, each point of the grid corresponds to 50 simulations with 1000 agents. One can identify four regions with different average $y$ values corresponding to the three dynamical regimes: two white zones (one on the left and the other one starting in diagonal from the lower middle part) corresponding to central convergence, one gray zone (drawing a triangle in the middle zone) for double extreme convergence and one dark gray zone (at the bottom right): single extreme convergence.
The dynamical regimes diagrams of Fig. 2 display large regions of intermediate $y$ values: "pure" dynamical regimes, corresponding to $y = 0$ or $0.5$ or $1$, are separated by "crossover" regions where intermediate average $y$ values and high standard deviation is due to a bimodal distribution of "pure" attractors depending on random sampling of initial conditions and pairing as confirmed by more detailed exploration \cite{Deffuantetal2002}.

\section{Opinion dynamics on networks}
\subsection{Social network as a regular grid}

Instead of a totally connected social network, we now run the model on a regular lattice (a torus) with a Moore neighborhood (connectivity $k = 8$). We systematically explore the parameter space for the uncertainty of the moderate individuals $U$ and the proportion of extremists $p_e$ (see Fig. 3).

\begin{center}
Fig.3
\end{center}

We note that:
\begin{itemize}
\item   $y$ is always below 0.6 which shows that the single extreme convergence never occurs. We only observed single extreme convergence in very particular cases of extremists positioning or for high values of $\delta$ (difference between initial proportions of extremists at each extreme). When the initial number of extremist is the same at each extreme, we only observe a higher final proportion of one extreme, but never the single extreme convergence observed in the fully connected case.
\item The transition from central to both extreme convergences has a shape similar to the one observed when the population is fully connected.
\end{itemize}

These observations can be explained as follows:
\begin{itemize}
\item For small values of $U$ a large number of clusters appear because the agents tend to be isolated: there is a high probability all their neighbors have too far opinions to be influential or to be influenced. This is also the case of the extremists, which are therefore not particularly influential for small $U$.
\item   For high values of $U$, the agents are on the contrary very likely to find interlocutors within their neighborhood. The influence of extremists propagates following the graph, first by attracting their own neighbors and then the others. The contamination is stopped when the formed cluster encounters another cluster of opposite opinion. Then the diffusion simply stops to invade the population.
\item   A possible explanation for the lack of single extreme convergence is due to this local propagation of extremism. It does not occur when the connectivity is high, because the majority always attracts back the agents which are occasionally attracted by one extreme (leading to the phenomenon of global drift to one extreme, when the majority losses contact with the other extreme). With the Moore neighborhood, for high values of $U$, each extremist disseminates very rapidly within its neighborhood, which prevents the single extreme convergence to appear.
\end{itemize}

If a De Moore neighborhood with an average connectivity of 8 is not sufficient to observe the single extreme convergence case, we have to increase this average connectivity. We then need a network model that enables to tune easily the average connectivity of the graph and also the network regularity. We selected the $\beta$-model of small-worlds by \cite{WattsandStrogatz1998} which satisfies these requirements.

\subsection{Exploration on small-worlds}
The $\beta$-model of small-world network is ruled by two parameters: the average connectivity $k$ and the randomness of the connections $p$. Starting from a regular structure (in our case a regular network over a circle) of connectivity $k$, we remove each link with the probability $p$, reconnecting it at random. In our case $k$ has to be odd because each individual on the circle has $k/2$ connections in each side.
We then studied the behavior of the model for a large uncertainty of the population $U=1.8$, and a low rate of extremist $p_e=0.05$ (bottom right part of the Fig. 2), leading to a single extreme convergence when the population is fully connected. We run 50 replications of the simulations on networks obtained when $p$ ranges from 0 to 1, thus from regular networks to totally random ones and $k$ ranges from 2 to 256 following the powers of 2 thus from 0.2\% to 25\% of the population (1000 individuals). Beyond the connectivity of 25\%, the behavior of the model is the same as in the totally connected case, whatever the value of $p$.

\begin{center}
Fig.4
\end{center}

We observe (see Fig. 4) a transition from double extreme convergence to single extreme convergence case when the connectivity ($k$) increases. Within the transition zone, the high standard deviation of $y$ indicates that sometimes a central convergence and single extreme convergence. The analysis of the traces of the opinions evolution for several simulations confirms the hypothesis expressed in section 3.a. For low connectivity each extremist influences its neighborhood which rapidly becomes extremist as well. We therefore obtain several clusters each one being in general controlled by one extremist, leading to a double extreme convergence. When the connectivity reaches a critical value, the population tends to regroup at the center and the fluctuations of this opinion cluster may lead to disconnect it from one extreme, leading to a drift to the other extreme. The central convergence takes place when the central majority looses contact with both extremes. This situation is favored by a low level of connectivity. When we increase the connectivity, this situation of a single extreme convergence regularly takes place as confirmed by the analysis of the distribution of convergence cases for a fixed value of $p$ (cf. Fig. 5).

\begin{center}
Fig.5
\end{center}

Moreover, the transition takes place for higher connectivity when $p$ decreases, i.e. when networks get more regular. Our hypothesis is that the regularity of the network reinforces the local effect which favors a fast local propagation of each initial extremist influence, leading to a double extreme convergence.
These results are robust considering the chosen substrate to build the network as for a square lattice Fig. 6 exhibits the same phenomenon.

\begin{center}
Fig.6
\end{center}

Then, we selected three typical couples ($U$, $p_e$) corresponding to different convergence types when the population is fully connected.

\begin{center}
Fig.7
\end{center}

All the chosen couples lead to quite similar dynamics when $p$ and
$k$ vary. For small connectivity we observe a majority of double
extreme convergences as for high connectivity we observe the
convergence case observed in the totally connected case. We
observe the systematic occurrence of central convergence in the
transition between these behaviors. Moreover, when we increase
$p$, we are closer to random networks and the transition occurs
for lower connectivity values. In Figure 7.b, there is a
transition zone between two zones of both extreme convergences.
The two observed both extreme convergences are due to different
processes. For low connectivity it results mainly from the
aggregation of local processes of convergence towards one extreme
locally and for higher connectivity, it results from a global
convergence of the central cluster, which is cut into two, each
part being attracted by an extreme.

\section{Discussion and perspectives}

We run the relative agreement model with extremists on small-world social networks, with varying connectivity and randomness. We found that there is a critical level of connectivity, which allows the single extreme convergence to take place. This critical level of connectivity increases when the regularity of the network increases. This result can be explained by the need of a first phase of global central clustering, to allow the single extreme convergence to take place. A low connectivity and high regularity of the network favor a fast local propagation of extremism, which prevents the global central clustering to take place. Comparing it to existing results \cite{BarratWeigt2000} we do not observe a classical small-world effect for networks that are at the same time highly clustered and that have a low diameter. But we found for our model that two phase transitions was observed in parameter zones that are usually kept unexplored, especially for high average degrees.
Other studies have to be conducted to understand the role of other parameters of the model. First simulations let us think that the population size is a critical parameter when the networks are regular.  The critique of these results in a sociological perspective is also a major challenge which is out of the scope of this paper. However, the idea of the necessity of a critical level of connectivity and some disorder in the network for extreme opinions to invade a population does not seem counterintuitive, and could find rich interpretation in real life sociological phenomena.
Moreover, several extensions of the model are under consideration. Among them, the addition of an opinion radicalisation dynamics that could result in a bigger gap between opinions after the interaction, is currently studied with a socio-psychologist.

\paragraph{Acknowledgments}
The authors are grateful to G\'erard Weisbuch and to the anonymous referees for their very valuable comments on this paper.
The work has been partially funded by the Conseil R\'egional d'Auvergne.

\begin{figure}
\begin{center}
\scalebox{0.5}{\includegraphics{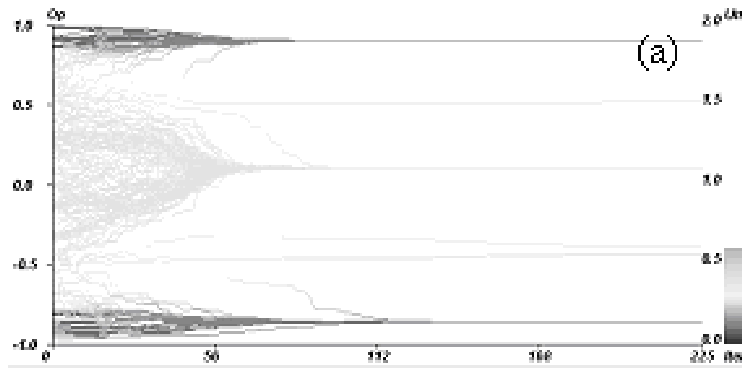}}
\scalebox{0.5}{\includegraphics{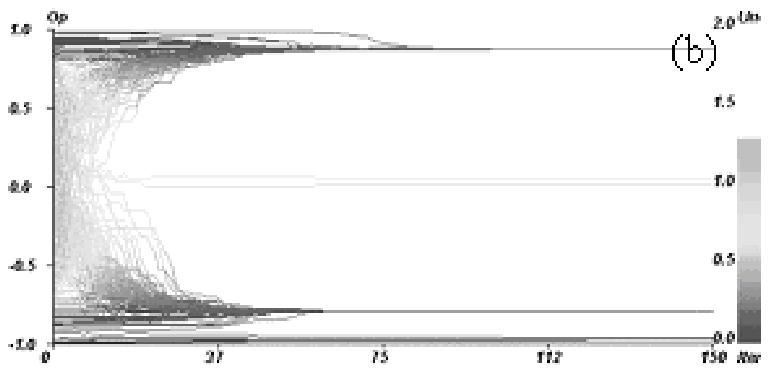}}

\scalebox{0.5}{\includegraphics{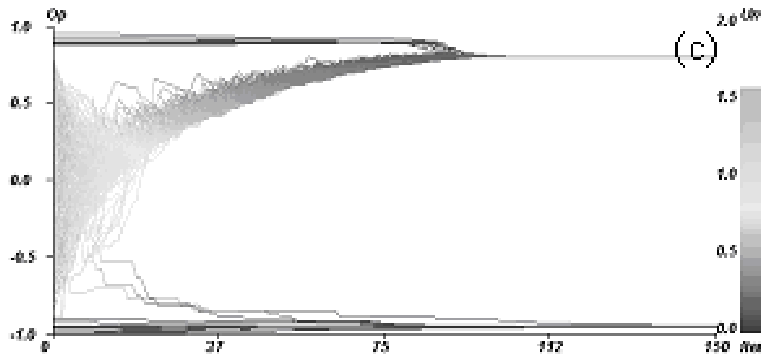}}
\scalebox{0.5}{\includegraphics{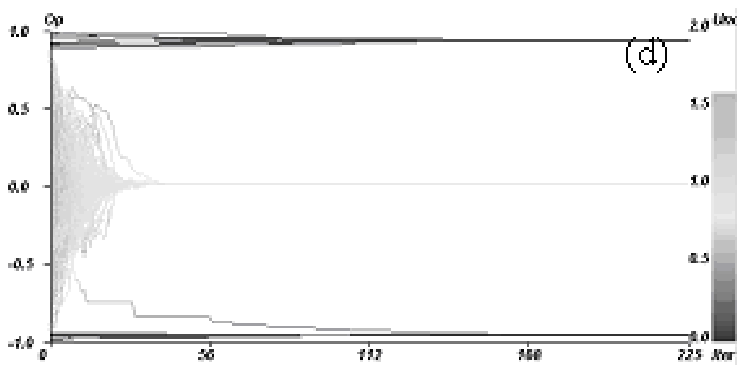}}
\end{center}
\caption{(a) Example of central convergence. Horizontal axis:
iterations. Vertical axis: opinions. Colored axis: uncertainties.
$p_e = 0.2, U = 0.4,  \mu = 0.5, \delta = 0, u_e = 0.1, N = 200$.
The majority (96\%) of the moderate agents are not attracted by
the extremes. (b) Example of both extremes convergence. $p_e =
0.25, U = 1.2, \mu = 0.5, \delta = 0, u_e = 0.1, N = 200$. The
initially moderate agents split and become extremists (43\% on the
positive side, 56\% on the negative side). (c) Example of single
extreme convergence. $p_e = 0.1, U = 1.4, \mu = 0.5, \delta = 0,
u_e = 0.1, N = 200$.  The majority (98.33\%) of initially moderate
agents is attracted by the negative extreme. (d) Central
convergence for the same parameters as in (c), the majority stays
at the center (Only one agent joins the negative extreme).}
\end{figure}

\begin{figure}
\begin{center}
\scalebox{0.5}{\includegraphics{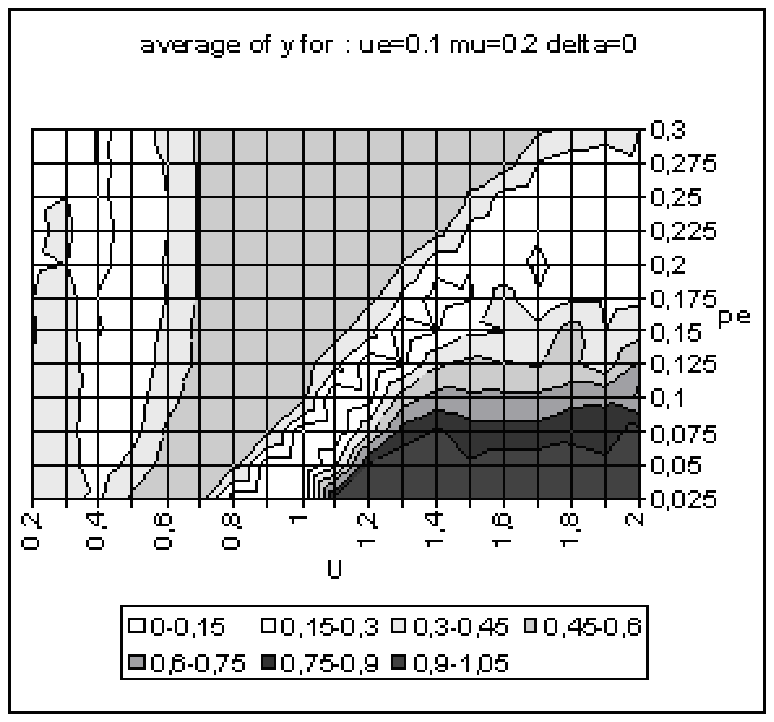}}
\scalebox{0.5}{\includegraphics{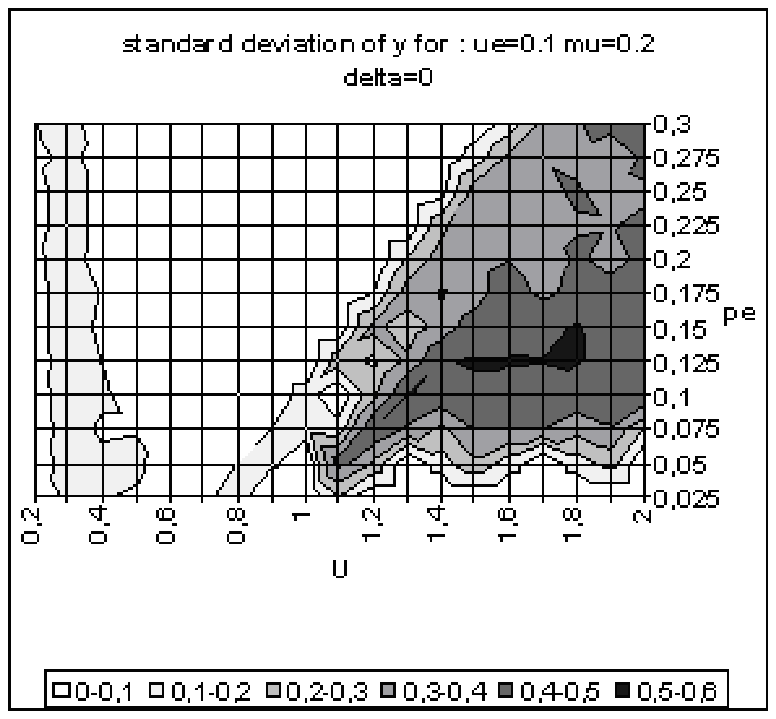}}
\scalebox{0.5}{\includegraphics{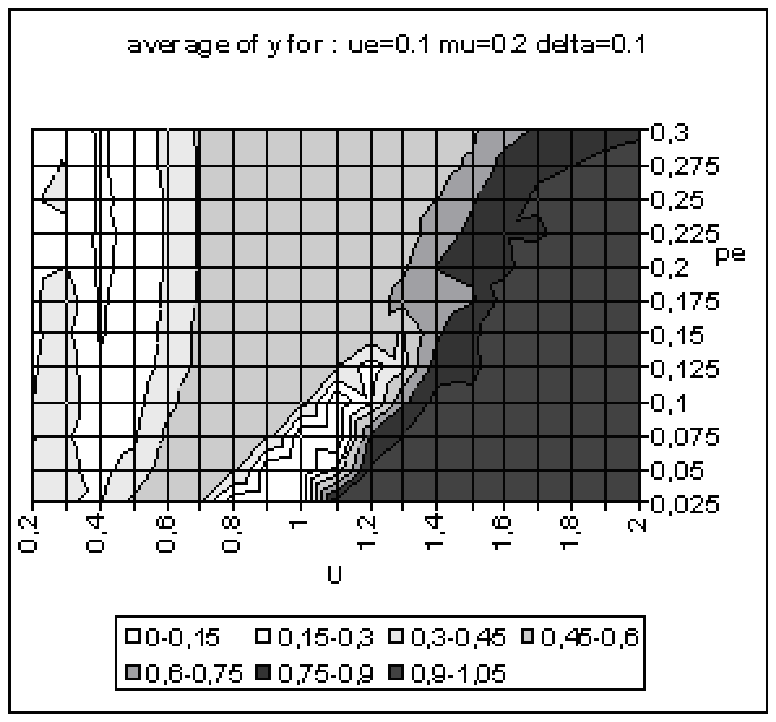}}
\scalebox{0.5}{\includegraphics{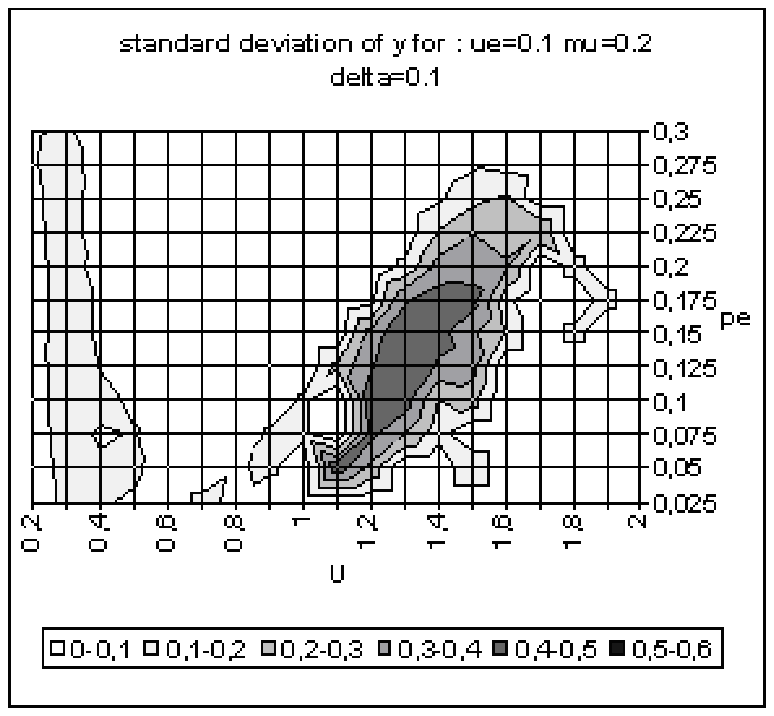}}
\end{center}
\caption{Typical pattern of average and standard deviation of
indicator $y$ (50 simulations at each point of the graph) as a
function of the uncertainty of the moderate agents ($U$) and the
global proportion of extremists ($p_e$) for $\delta = 0$ (top) and
$\delta = 0.1$ (bottom). The other parameters are fixed:
uncertainty of the extremists $u_e = 0.1$, intensity of
interactions $\mu = 0.2$, initial relative difference between the
extremists, $\delta = 0.1$. On the graph of average $y$, the
yellow or white zones on the left part correspond to central
convergence, the orange, typically in the upper middle part to
both extremes, and brown at the bottom right to single extreme.}
\end{figure}

\begin{figure}
\begin{center}
\scalebox{0.8}{\includegraphics{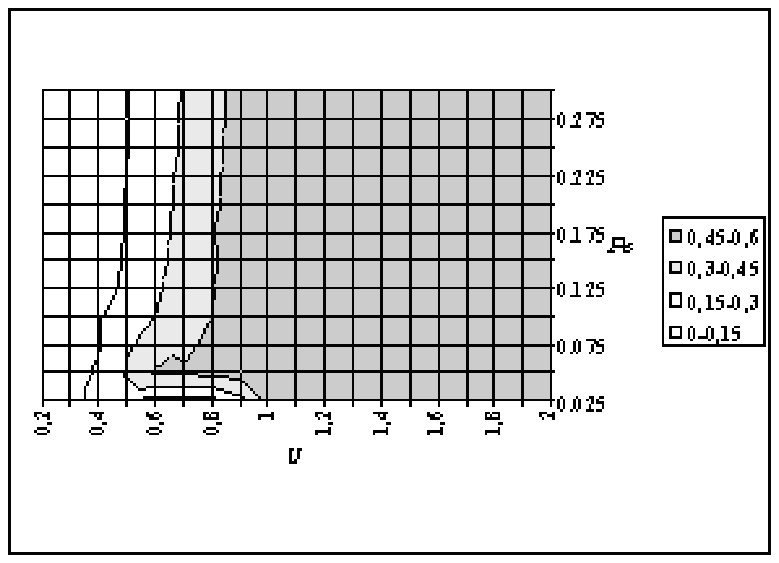}}
\scalebox{0.5}{\includegraphics{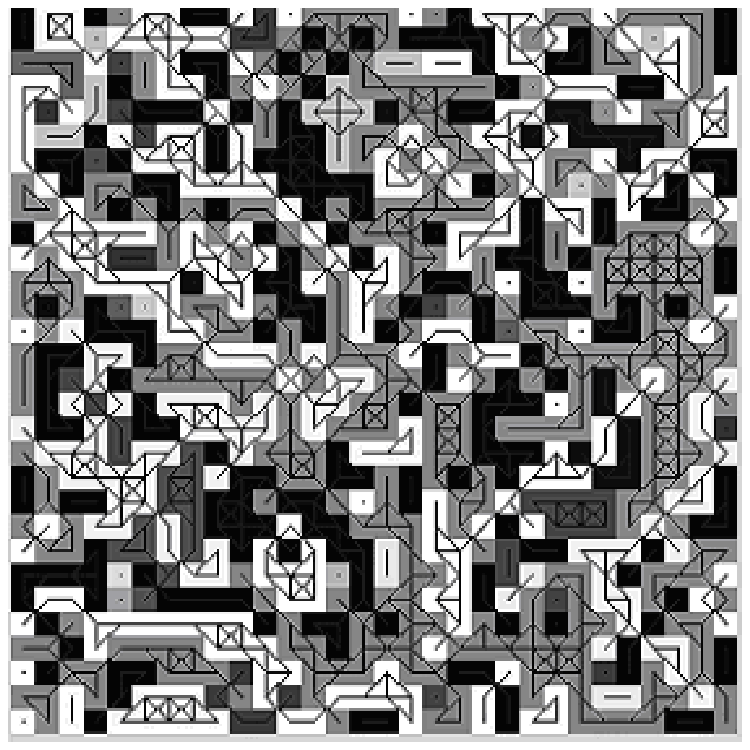}}
\scalebox{0.5}{\includegraphics{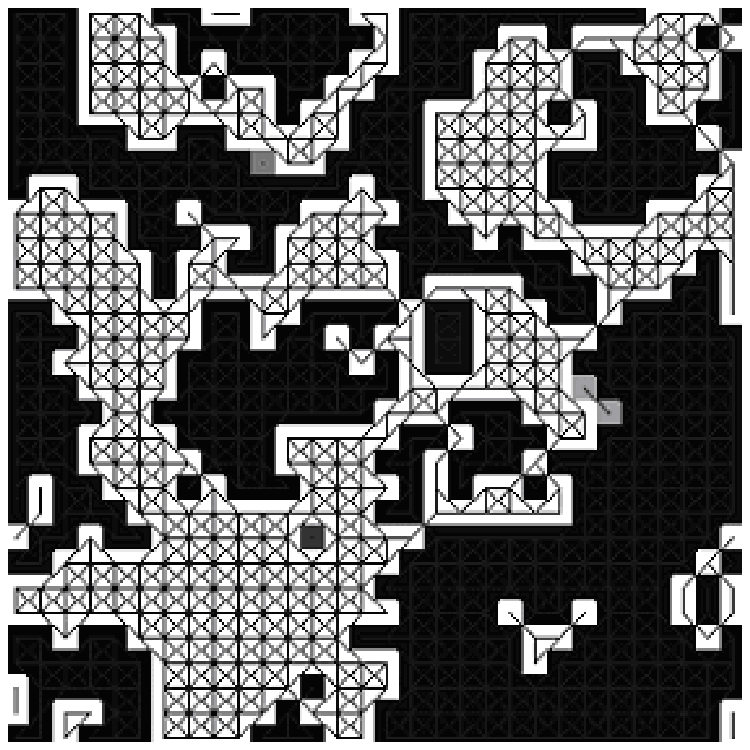}}
\end{center}
\caption{(a) Exploration of the parameter space on a grid formed
by $U$ and $p_e$, with a Moore's neighborhood (connectivity 8),
other parameters are $u_e=0.1$ and $\mu =0.2, \delta =0$. The
associated standard deviation is relatively low except for the
zone corresponding to the transition for low values of $p_e$. The
main observations are the general decrease of the $y$ values and
the absence of single extreme convergence ($y$ never reaches
values close to 1). (b) Final state of the system corresponding to
the left part of a) for $U=0.4$ and $p_e=0.2$, the system is
highly clustered. (c) Final state corresponding to the right part
for $U=1.4$ and $p_e=0.2$, the system, even converging towards a
double extreme, is much more cohesive.}
\end{figure}

\begin{figure}
\begin{center}
\scalebox{0.6}{\includegraphics{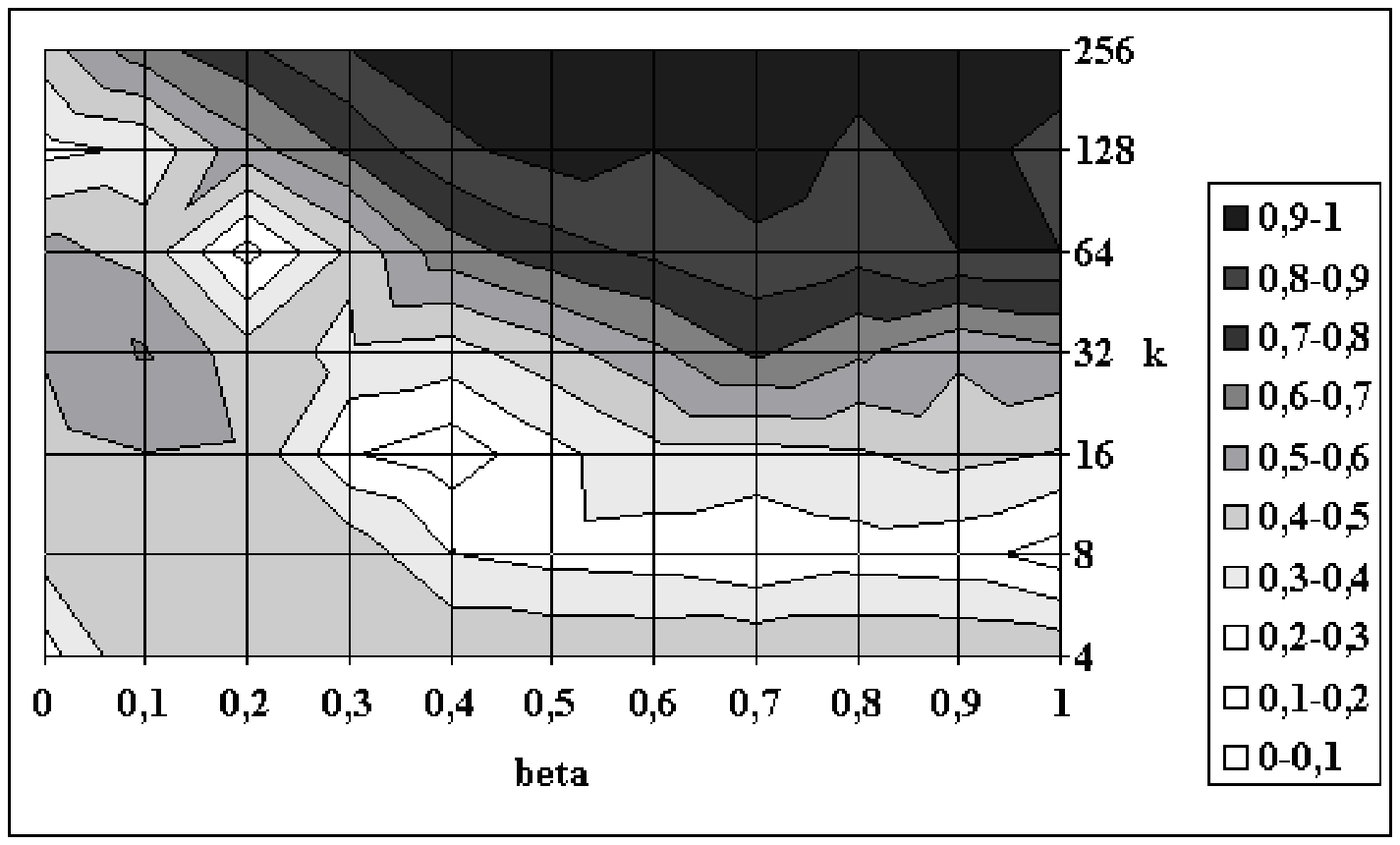}}
\scalebox{0.6}{\includegraphics{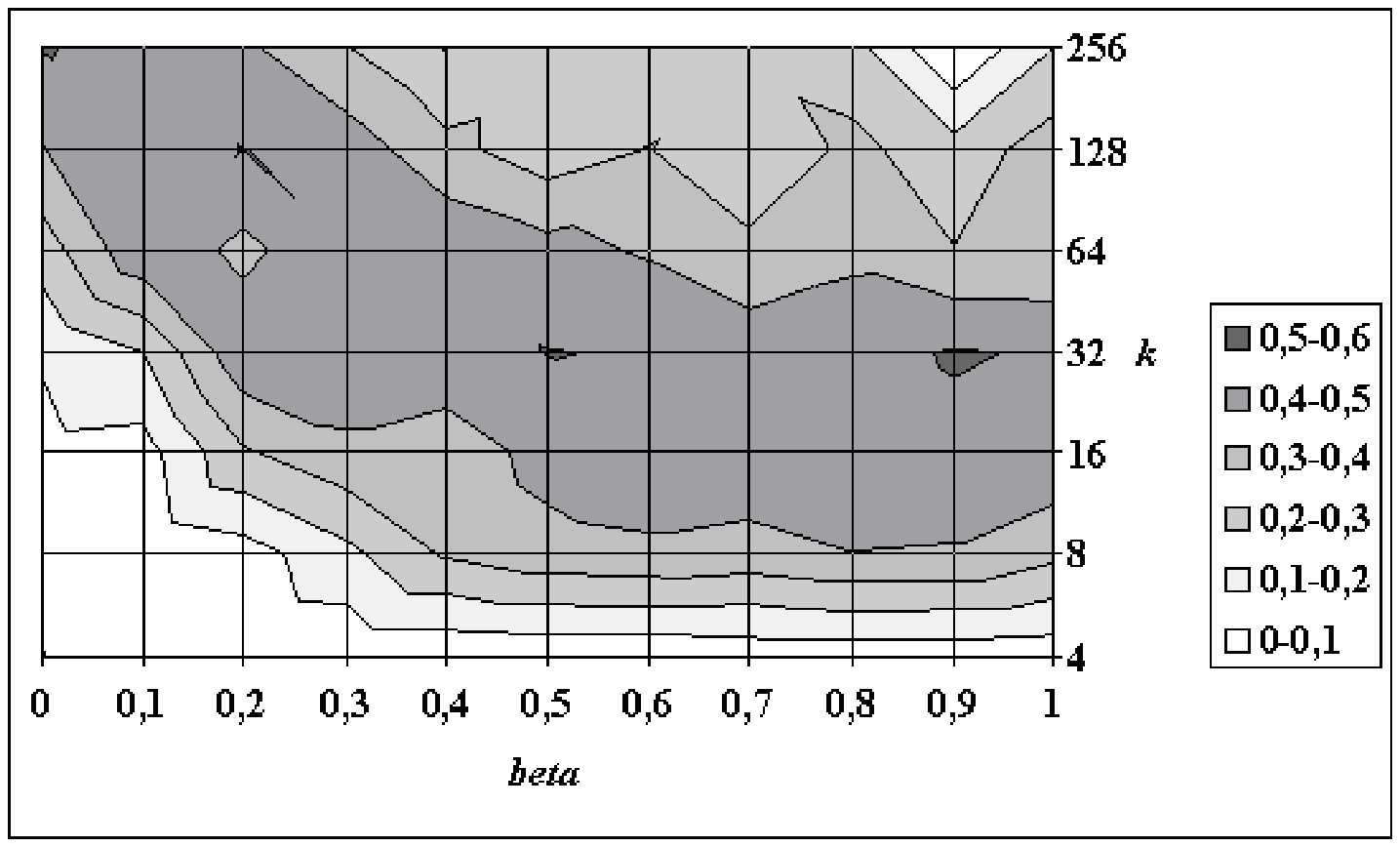}}
\end{center}
\caption{Exploration of the effect of a network following a
small-world topology with a ring substrate ($\beta$ and $k$
connectivity parameter) on the dynamics of the model for $U=1.8,
u_e=0.1, N=1000, \mu=0.1, \delta=0, p_e= 0.05$ (single extreme
convergence for the totally connected case)}
\end{figure}

\begin{figure}
\begin{center}
\scalebox{0.45}{\includegraphics{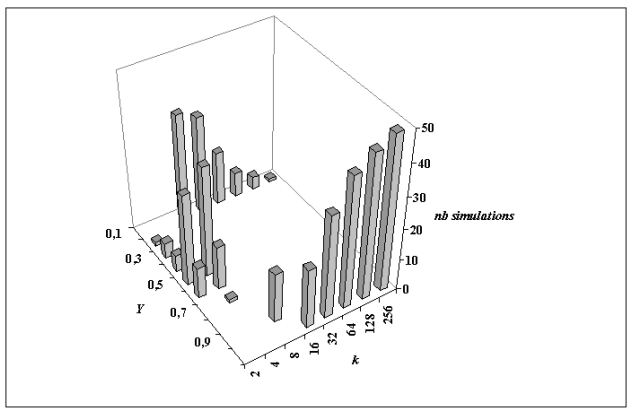}}
\end{center}
\caption{Distribution of $y$ for a small-world topology with $k$
connectivity parameter on a logarithmic scale, for $\beta = 0.8,
U=1.8, u_e=0.1, N=1000, \mu=0.1, \delta=0, p_e= 0.05$. We observe
the phase transition that occurs for values of connectivity around
8 and from that point that the mixing between single convergence
case and central convergence tends to become uniformly single
extreme convergence.}
\end{figure}

\begin{figure}
\begin{center}
\scalebox{0.3}{\includegraphics{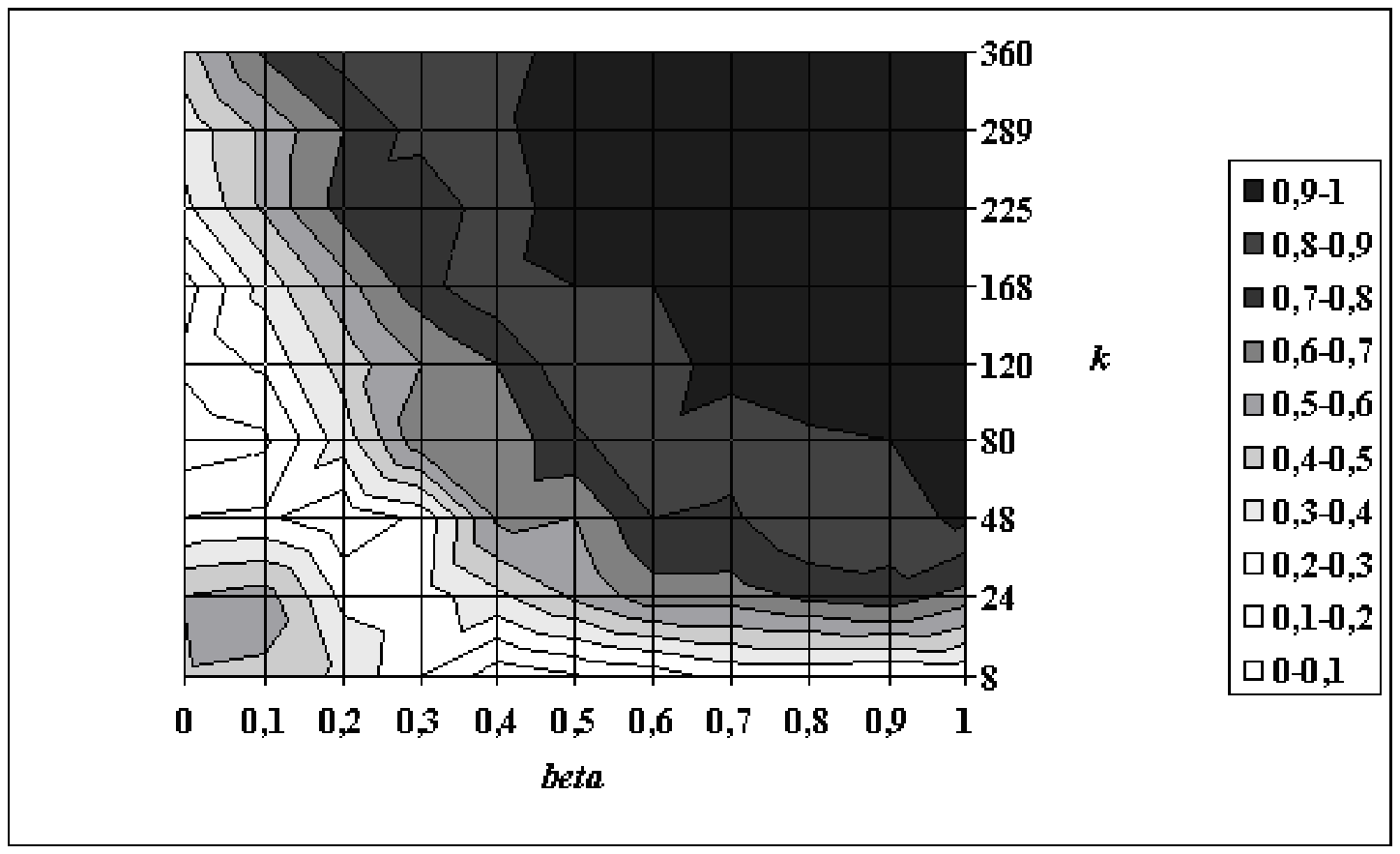}}
\scalebox{0.3}{\includegraphics{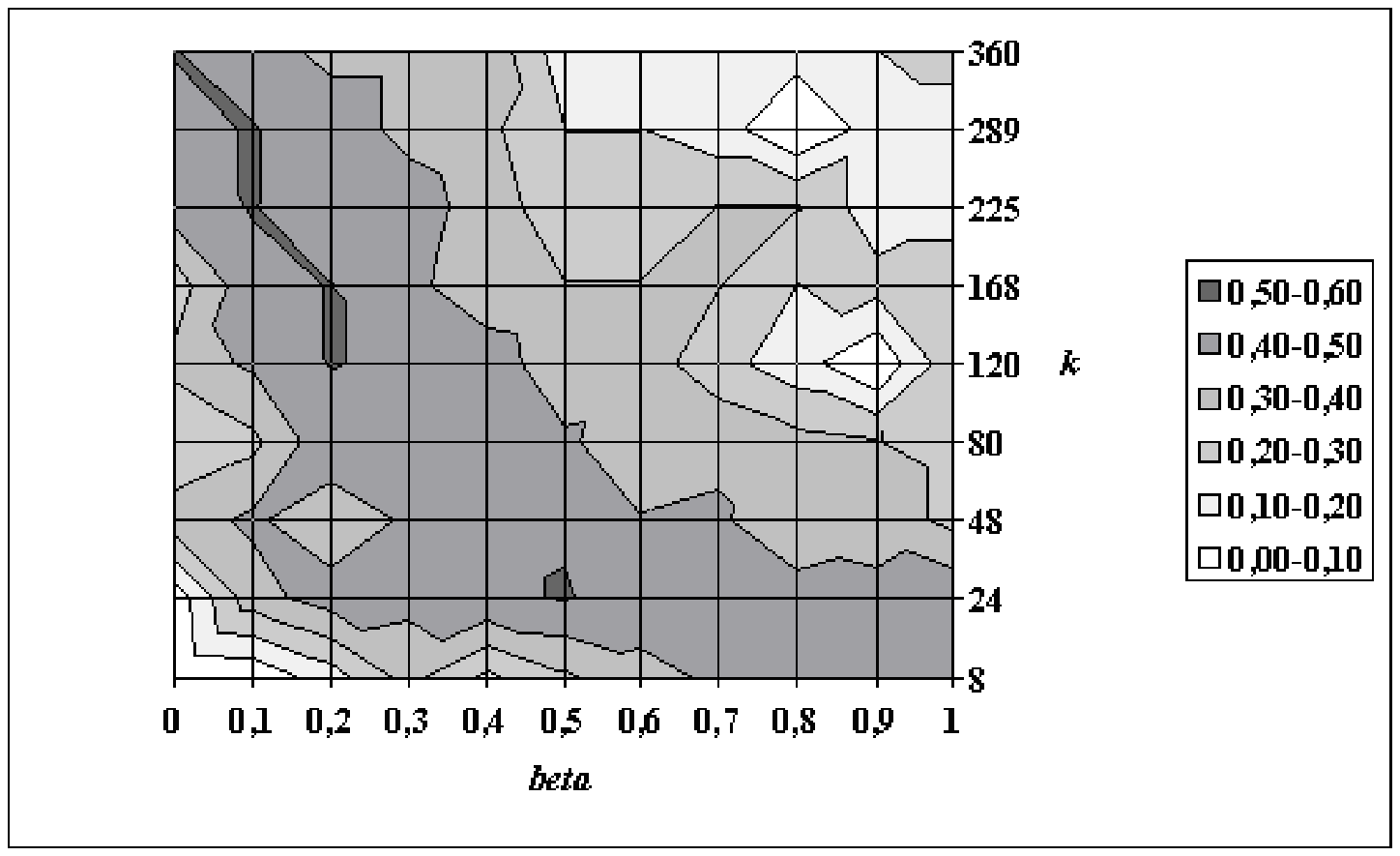}}
\end{center}
\caption{Exploration of the effect of a network following a
small-world topology with a grid substrate applying a generalized
Moore neighborhood ($\beta$ and $k$ connectivity parameter) on the
dynamics of the model for $U=1.8, u_e=0.1, N=1000, \mu=0.1,
\delta=0, p_e= 0.05$ (single extreme convergence for the totally
connected case). Qualitative results are the same compared to a
ring substrate even distortions are observed due to possible
connectivity values to be tested.}
\end{figure}

\begin{figure}
\begin{center}
\scalebox{0.3}{\includegraphics{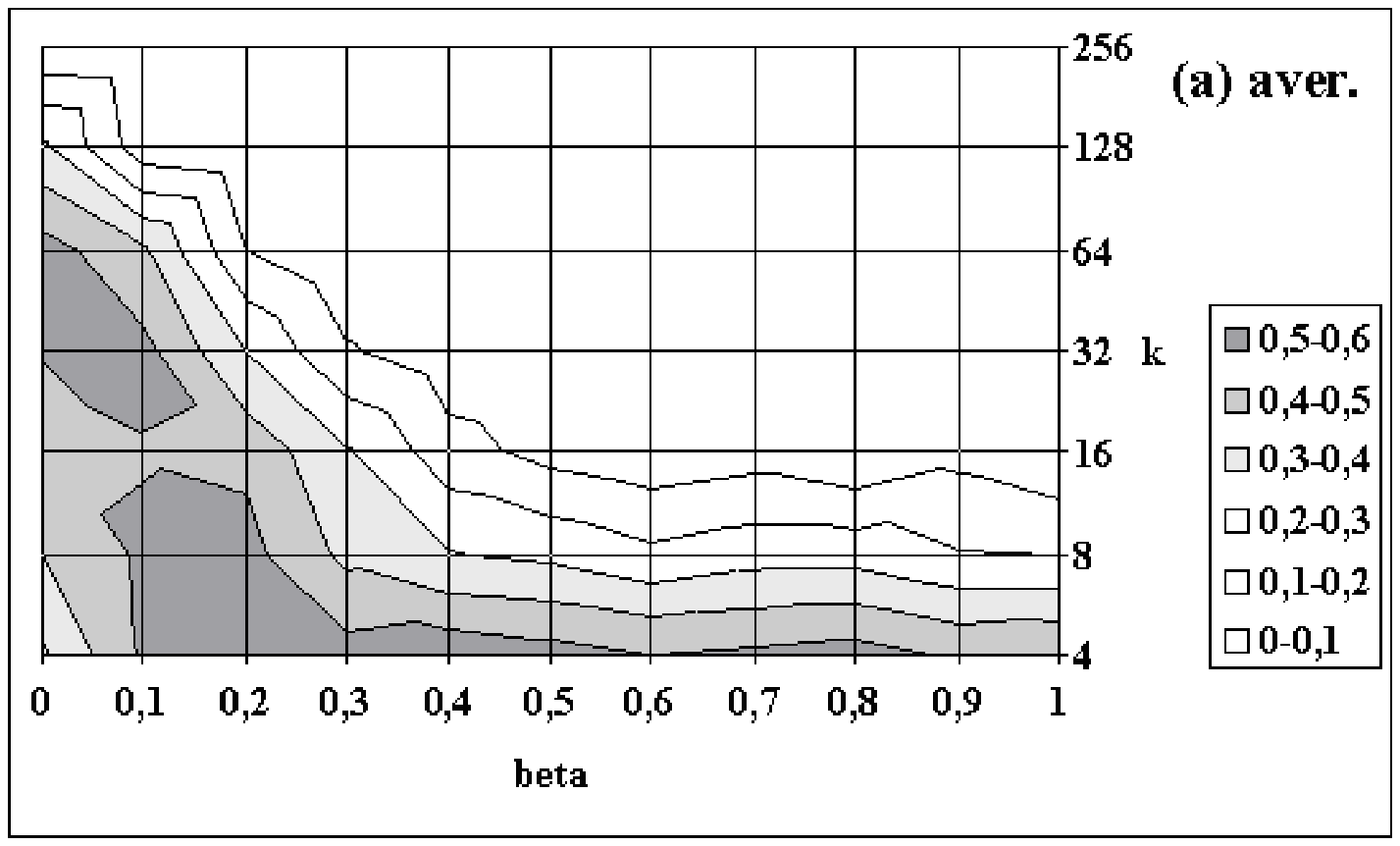}}
\scalebox{0.3}{\includegraphics{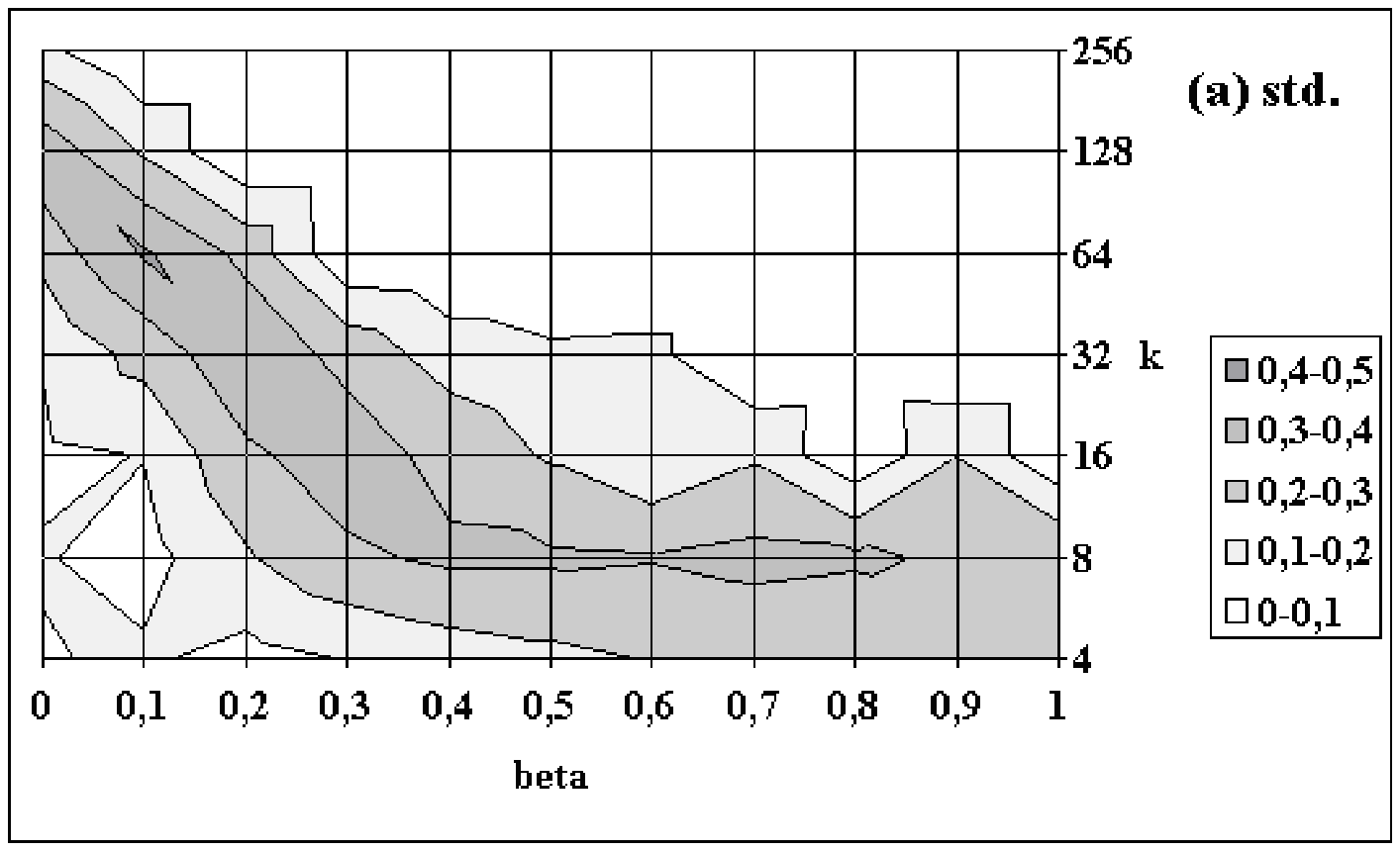}}
\scalebox{0.3}{\includegraphics{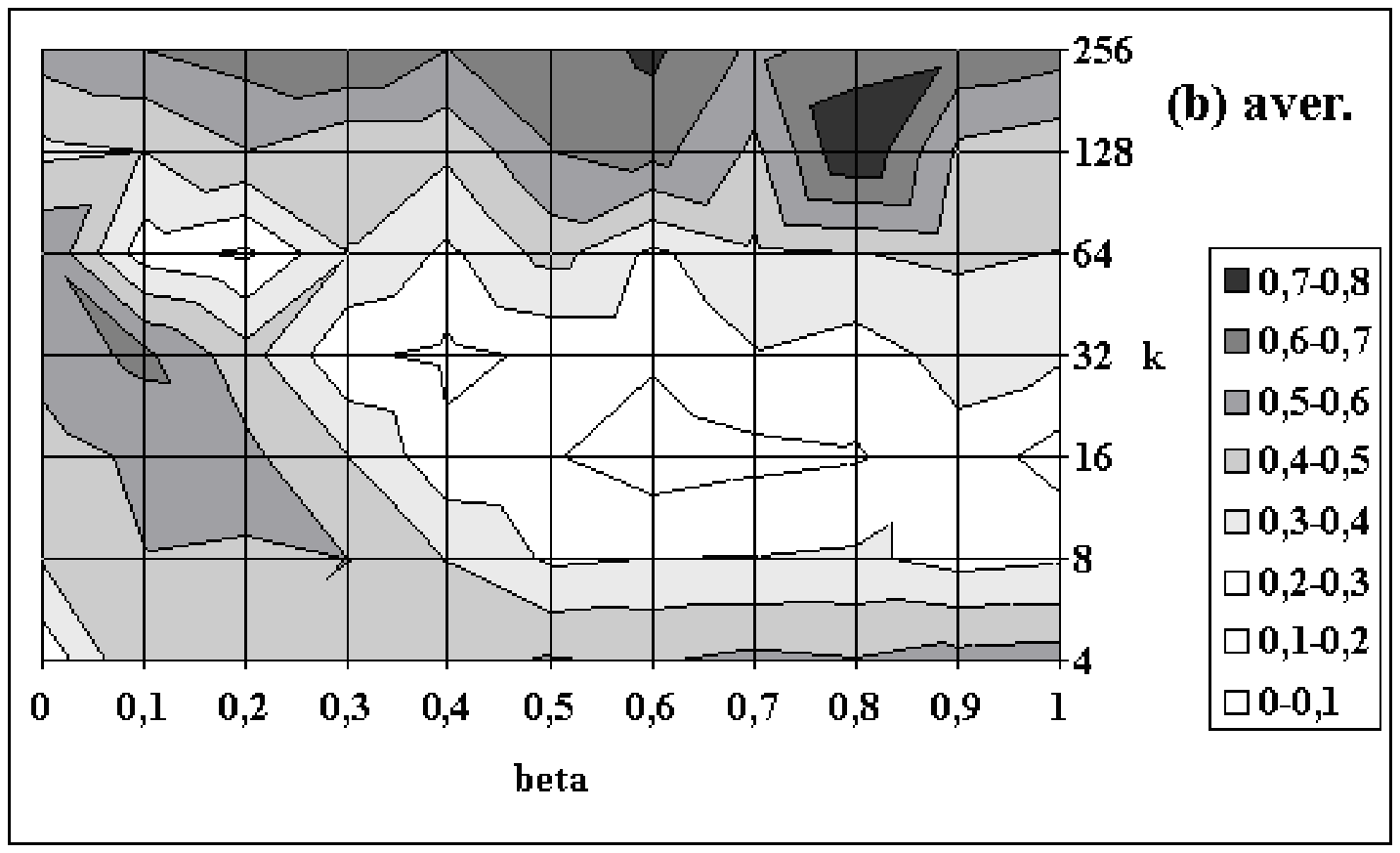}}
\scalebox{0.3}{\includegraphics{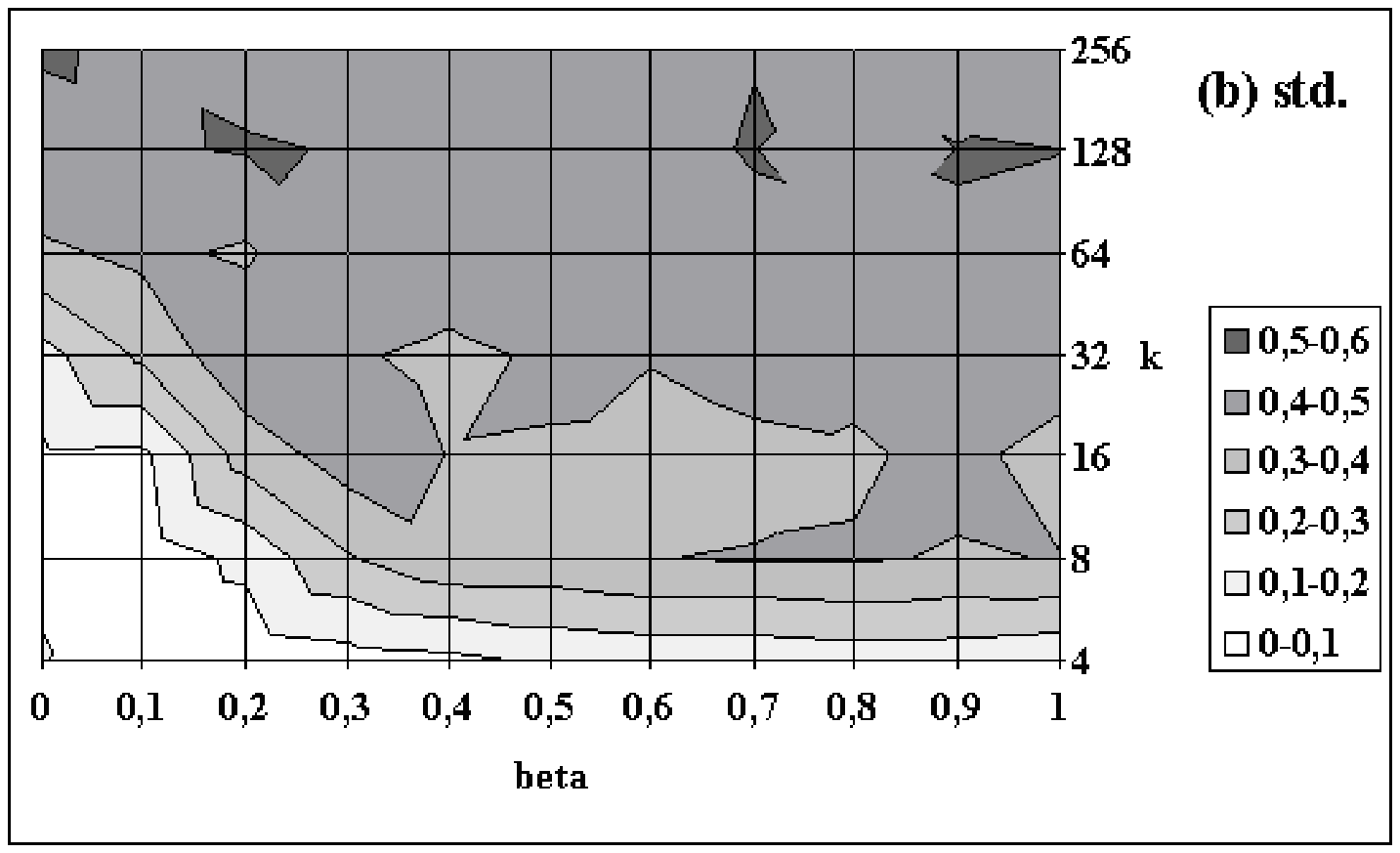}}
\scalebox{0.3}{\includegraphics{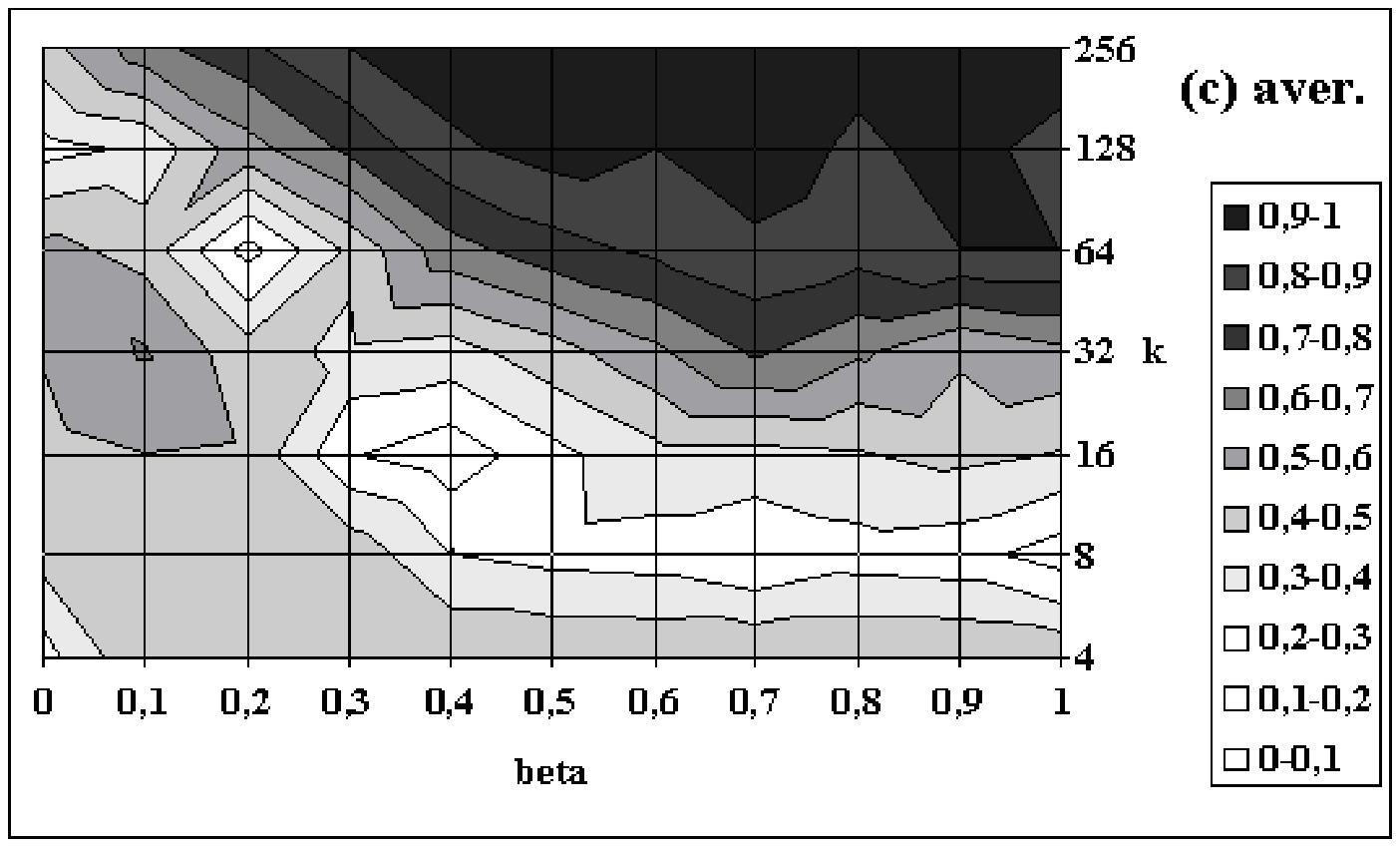}}
\scalebox{0.3}{\includegraphics{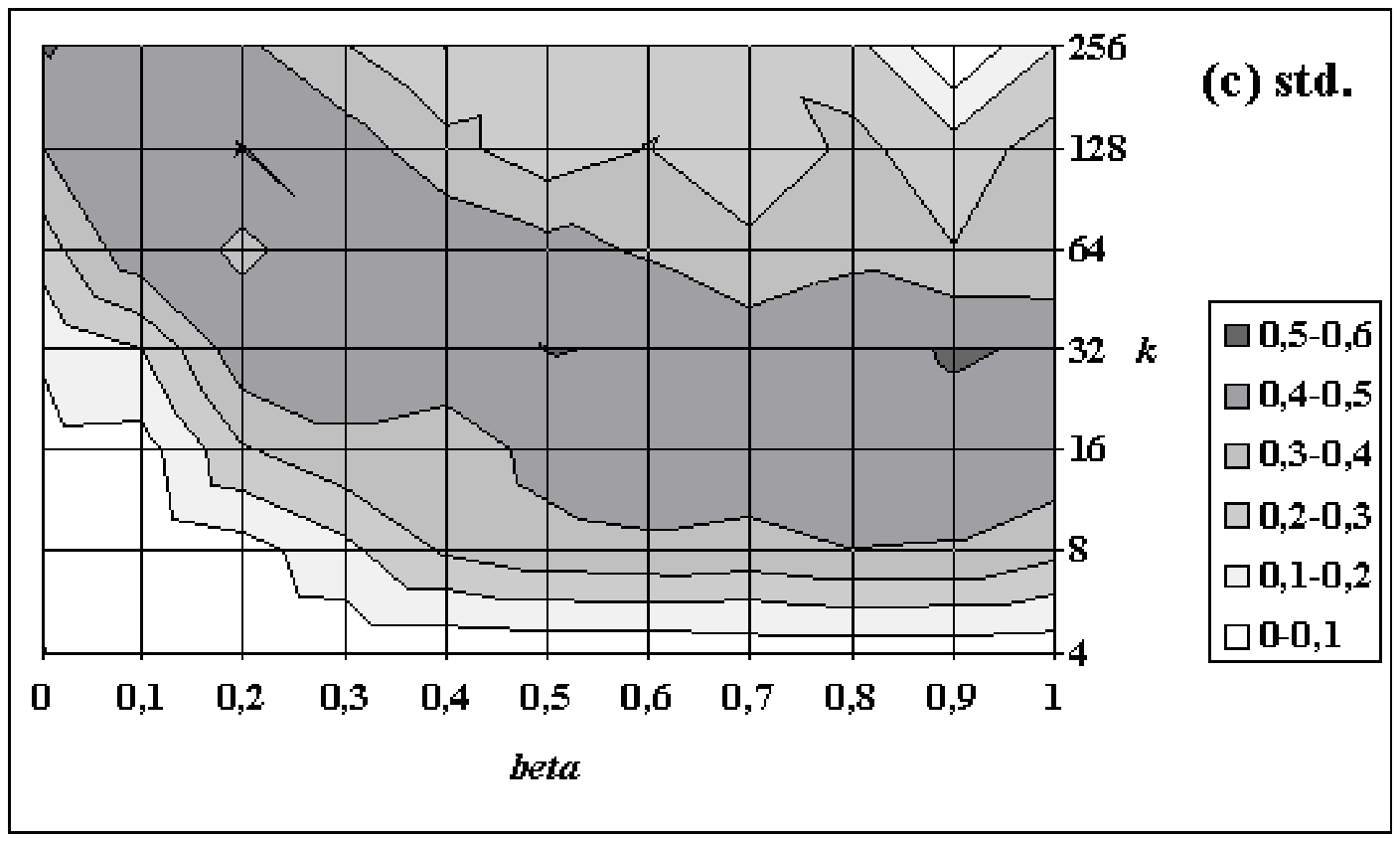}}
\end{center}
\caption{Representation of the average of $y$ over 50 replications
at each point on the left and standard deviations on the right for
different couples ($U, p_e$) other parameters are taken constant
considering the Fig. 4. The three points taken corresponds in the
totally connected case to (a) central convergence case ($U=1.0$
and $p_e= 0.05$) (b) double extreme convergence case ($U=1.2$ and
$p_e=0.05$) and (c) single extreme convergence case ($U=1.4$ and
$p_e=0.05$) as we can observe when we increase the connectivity.}
\end{figure}


\begin{thebibliography}{  }

\bibitem[1]{Deffuantetal2002} G. Deffuant, F. Amblard, G. Weisbuch, T. Faure, J. Artificial Societies and Social Simulation, 5 (2002) 4.
\bibitem[2]{WattsandStrogatz1998} D.J. Watts, S.H. Strogatz, Nature 393 (1998) 440.
\bibitem[3]{BarabasiandAlbert2002} A.-L. Barabasi, R. Albert, Rev. Mod. Phys. 74 (2002) 47.
\bibitem[4]{Barabasi2002} A.-L. Barabasi, Linked : The New Science of Networks, Perseus Publishing, (2002).
\bibitem[5]{WeidlichHagg1983} W. Weidlich and G. Haag, Concepts and models of quantitative sociology (Springer, Berlin, 1983).
\bibitem[6]{Weidlich1991} W. Weidlich, Phys. Rep. 204 (1991) 1.
\bibitem[7]{Galam1997} S. Galam, Physica A 238 (1997) 66-80.
\bibitem[8]{Sznajd2000} K. Sznajd-Weron and J. Sznajd, Int. J. Mod. Phys. C 11, 1157 (2000); D. Stauffer, A.O. Sousa and S. Moss de Oliveira, Int. J. Mod. Phys. C 11, 1239 (2000); K. Sznajd-Weron and R. Weron, Int. J. Mod. Phys. C 13, No. 1 (2002); A.S. Elgazzar, Int. J. Mod. Phys. C 12, No. 10 (2001).
\bibitem[9]{BarratWeigt2000} A. Barrat, M. Weigt, Eur. Phys. J. B 13, 547 (2000).
\bibitem[10]{KupermanZanette2002} M. Kuperman, D. Zanette, Eur. Phys. J. B 26, 387 (2002).
\bibitem[11]{HegselmannetKrause2002} R. Hegselmann, U. Krause, U., J. Artificial Societies and Social Simulation, 5 (2002) 3.
\bibitem[12]{Deffuantetal2001} G. Deffuant, D. Neau, F. Amblard, G. Weisbuch, Adv. Complex Systems, 3 (2001) 87.
\bibitem[13]{Weisbuchetal2002a} G. Weisbuch, G. Deffuant, F. Amblard, J.-P. Nadal, Complexity, 7 (2002) 55.
\bibitem[14]{Weisbuchetal2002b} G. Weisbuch, G. Deffuant, F. Amblard, J.-P. Nadal, Lect. Notes Econ. Math. Sys., 521 (2002) 225.
\bibitem[15]{Axelrod1997} R. Axelrod, The complexity of cooperation (Princeton University Press, Princeton, NJ, 1997).
\bibitem[16]{Faureetal2003} T. Faure, Physica A, submitted.
\end{thebibliography}
\end{document}